\documentclass[aps,twocolumn,groupedaddress,showpacs,floatfix,notitlepage,color]{revtex4-1}
\usepackage{graphicx}
\usepackage{epsfig}
\usepackage{amsfonts}
\usepackage{color}
\usepackage{ulem}
\usepackage{amsmath,amssymb}
\usepackage[autostyle]{csquotes}
\usepackage{hyperref}
\usepackage[percent]{overpic}
\usepackage{pict2e}

  \makeatletter
    \renewcommand\@make@capt@title[2]{%
     \@ifx@empty\float@link{\@firstofone}{\expandafter\href\expandafter{\float@link}}%
      {\textbf{#1}}\@caption@fignum@sep#2\quad}%
    \makeatother
   
\makeatletter
\renewcommand{\fnum@figure}{\textbf{Figure~\thefigure}}
\makeatother

\begin{document}

\title{Poverty Index  With Time Varying Consumption and Income Distributions}

\author{
Amit K Chattopadhyay}
\affiliation{Aston University, Mathematics, Birmingham B4 7ET, UK}
\email[Corresponding author: ] {a.k.chattopadhyay@aston.ac.uk}

\author{T Krishna Kumar}
\affiliation{Rockville-Analytics, Rockville, MD, USA}
\email{tkkumar@gmail.com}

\author{Sushanta K Mallick}
\affiliation{School of Business and Management, Queen Mary University of London, Mile End Road, London E1 4NS, UK}
\email{s.k.mallick@qmul.ac.uk}

\begin{abstract}
In a recent work (Chattopadhyay, A. K. et al, Europhys. Lett. {\bf 91}, 58003, 2010), based on food consumption statistics, we showed how a stochastic agent based model could represent the time variation of the income distribution statistics in a developing economy, thereby leading to the definition of an alternative and more preferable \enquote{poverty index} (PI) that compared favorably with the actual poverty gap (PG) index data. The new poverty index was defined based on two variables, the probability density function (PDF) of the income statistics and the consumption deprivation (CD) function, a mathematical function representing the shortfall in the minimum consumption needed for survival. This starting model used an agent-based interaction mechanism for asset transactions to generate the time-dependent income distribution, while leaving unexplained how the minimum consumption needed for survival is determined in a time dependent manner. The time dependence of the CD function was introduced there through data extrapolation only and not through endogenous generation of a time dependent series. The present article overcomes these limitations and arrives at a new probabilistic paradigm of poverty modeling. A new unified theoretical structure has been developed that treats all economic agents as interacting agents engaged in the trade of goods and services leading to time varying consumption and income distributions. Here the minimum threshold of income is defined by consumption deprivation (CD) where commodity trade would not take place beyond the basic necessities. Our results reveal that the nature of time variation of the CD function leads to a downward trend in the threshold level of consumption of basic necessities, suggesting a possible dietary transition in terms of lower saturation level of food-grain consumption required for survival. This study reinforces and strengthens our benchmark original study on poverty analysis based on Engel curves and consumption deprivation. The poverty index profile presented here, based on a dynamic model of agent-based trading, conforms to recently observed trends more closely than what the conventional measures of poverty (head count index, poverty gap index and squared poverty gap index) have failed to depict.  
\end{abstract}
\date{\today}
\maketitle

\section{Introduction}
\label{introduction}

Trade involves asset exchange between partners who agree to a bartering network between themselves. Within this network, there are pockets of individuals or families whose rate of flow of income derived from exchanging their own human skills in the market place is not adequate to meet their basic necessities. Such people are referred to as poor and their state of living or consumption is called poverty. In order to understand the mechanics of income and consumption of such individuals, and the associated measure of poverty we embed them within a broader community of individuals endowed both with human skills as well as accumulated wealth and place them in an interacting market environment and a non-market state.

\par
Starting with the famous work of Pareto \cite{Pareto_1895}, numerous (mostly time independent) studies \cite{Pareto_1895,Mandelbrot_1960,Montroll_1983,Bouchaud_book,Bouchaud_2008,Mantegna_1995,Podobnik_2012,Mohanty_2006}  have been made on the high income side, all generally converging to the same identical outcome, that of a power law tail for the large income sectors. Similar power law behaviors have been noted in other noisy models involving complex networks \cite{Carro_2016, Vespignani_2013}. The low income sectors, more specifically the \enquote{destitution margin}, has remained largely under studied.  
In this article, we will extend the understanding obtained from our previous stochastic agent-based model for generating a time varying profile of income distribution \cite{Chattopadhyay_2010}, and Engel curve based analysis of poverty \cite{Kakwani_1980, Sitaramam_1996, Kumar_1996}. 
In our first work \cite{Chattopadhyay_2010}, the time dependence of consumption was exogenously given (and hence could not be predicted). 
In this article, we generate time dependence of both income and consumption endogenously using the same agent-based trading model. 
In \cite{Sitaramam_1996} and \cite{Kumar_1996}, the underlying market mechanism for exchange of goods and services (commodities and assets) remained unexplained. In this paper, we apply the same agent-based commodity exchange model as the basis for generating the Engel curve. 
The most significant outcome of this marriage in concepts propounded in the older papers with the new one is that the characteristic features of poverty are endogenously demarcated by the nature of trade in an agent-based model. 
In particular, we find agents do not engage in trading of assets or even of luxuries as long as their basic necessities are not fulfilled, 
a feature emerging from the work by Engel, later studied by Kumar, Gore and Sitaramam \cite{Kumar_1996} and Chai and Moneta \cite{Engel_1895} and \cite{Chai_2010}. 
The poverty threshold thus emerges endogenously from the model as that level of income at which income is less than consumption deprivation of essential commodities. We will use, as our defining constituent of asset exchange, food grain consumption statistics  data availed from World Bank \cite{NSS}.

\par
As always, the paucity of basic resources leading to poverty can be attributed to inhomogeneity in the spread of such resources that can be largely remedied through decentralization of assets. The history of economics is replete with examples of attempts at appropriating the best possible scheme for effecting this decentralization mechanism. The issue is contentious and attempts have generally been fractious \cite{Kumar_2009} with political opinions governing the subset of assumptions based on which most of the theoretical models have been attempted. In this work, our premise will be dispassionate economic argument that is free of socio-political conduits and limit our analysis to data based facts, that will define the model largesse. This will thus be a bottom-up approach where a model will be constructed around known facts and then analyzed for quantitative agreement with real data.

\par
Unfortunately poverty studies in economics made a fictitious classification of all people into two categories called \enquote{poor} and \enquote{non-poor} by defining a cut-off point of poverty level income (the so called \enquote{focus axiom} by Sen \cite{Sen_1976}). Several controversial issues arise on how that poverty level income is determined and if one can really say that someone just above that line is so different from the one just below that level to be excluded from poverty studies, etc. Defining who the poor are and then to define poverty is like putting the cart before the horse. One must define what poverty is and then identify those who have high degrees of poverty and label them as poor. Poverty then can be defined as deprivation in consumption of essential commodities, such as a staple food, say cereals.  This is how Engel studied poverty. Some of our earlier works highlight these aspects in more details \cite{Kumar_2009, Kumar_1996, Sitaramam_1996}. The main thrust of Engel curve-based measure of poverty was to reject the \enquote{focus axiom} of Sen \cite{Sen_1976} and decline the notion of the existence of a fine distinction between the poor and non-poor. Although this argument was not well received by the mainstream economists when the seeds of the idea were initially advanced in \cite{Sitaramam_1996}, the arbitrary nature of the poverty line is now being addressed with renewed interest across the academic community.

\par
The starting point of our approach in this paper is to assume that there are two distinct markets, one for assets that are accumulated and saved or dissolved over time. The other market is the commodity market where goods and services needed for current consumption are traded. We assume that all people are permitted to participate in trade exchange over both markets, while a majority of the poor have a limited participation in the asset market transaction, they only trade their human skills or human capital for wage income, which is used in the other commodity markets for exchange. We do not introduce the notion of the poor and poverty line on our own, exogenously. Instead we let the data tell us what kind of asset and commodity exchanges people are involved in, observe the equilibrium behavior in the commodity and asset markets to determine income distribution and the implied consumption distribution, consumption deprivation and poverty.

\subsection{The Engel Profiles: Definition of Consumption-Deprivation}
\label{engel_plots}

\par
\begin{center}
\begin{figure}[tbp]
\includegraphics[height=8.0cm,width=9cm]{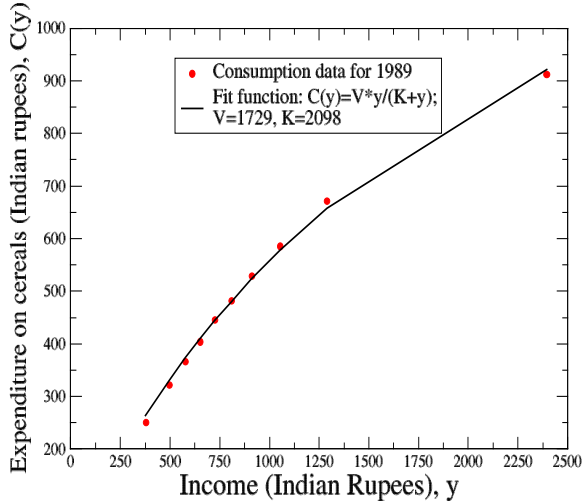}
\caption{A plot of the consumption function $C(y)$ against the expenditure $y$ on food grains. The data (dots) have been fitted against the consumption function $C(y)=\dfrac{Vy}{K+y}$ using a least squares mechanism.
\label{fig_consumption}}
\end{figure}
\end{center}

As showed in two of our previous works \cite{Chattopadhyay_2010,Chattopadhyay_2007}, all three forms of conventional poverty measures, \enquote{headcount index},\enquote{poverty gap index} and \enquote{square poverty gap index} suffer from an inherent arbitrariness in defining an appropriate "{poverty line}'' across which each such index may be measured. These are classed as statistical measures \cite{Kakwani_1980}. 
In those studies poverty measure was defined with an exogenously and arbitrarily defined poverty line to identify who the poor are and the data were used only to estimate poverty so defined. In this study, and some of our earlier studies \cite{Sitaramam_1996,Kumar_1996,Kumar_2009} based on Engel curves, we used data to reveal the pattern of consumption deprivation on essential commodities that gave us an observed description of existing poverty. Thus our approach had an inherent congruence between theory and empirical verification. To define a more robust poverty measure, as in \cite{Chattopadhyay_2010}, we use statistics for
consumption deprivation (CD) 
\cite{Kumar_1996,Sitaramam_1996}. CD is derived from consumption expressed as a function of income. As shown in Figure \ref{fig_consumption}, consumption shows an initial increase with income followed by saturation, reflecting the fact that one can not eat in proportion to an increase in earning \cite{Cirera_2010}. CD is a shortfall in consumption from this saturation level. The complementary picture is shown in Figure 2, that was previously derived and discussed in detail in \cite{Chattopadhyay_2010}; it shows CPI-normalized deflated statistics of the income variation of the Cumulative Distribution Functions (CDF) against CPI-normalized income $y$. The results showed a highly interesting data collapse over all years that suggested inherent scaling in the CDF statistics.

Correlating the NSS income data $y$ with that of the expenditure on basic cereals $C(y)$, we find a good fit to a functional relationship \cite {Chattopadhyay_2007}:

\begin{equation}
C(y,t)= \frac{V(t)y}{K(t)+y}
\label{C_equation}
\end{equation}

where the parameters $V(t)$ and $K(t)$ are time dependent \cite{Chattopadhyay_2007, Kumar_2009}, as shown in figures \ref{fig_Vplot} and \ref{fig_Kplot}. $V(t)$ is a measure of the overall amount spent per family on cereal consumption when income is very large. This is the value of $C(y)$ in the plateau region of Fig. \ref{fig_consumption} for $y \to \infty$. $K(t)$, on the other hand, can be taken as the income necessary to support a consumption expense that is half of the maximum overall spend per family ($V(t)$). Incidentally, $K$ is also that level of income where consumption deprivation function and the affluence function (the actual consumption) will be equal. Overall, the {\it consumption deprivation} relates to the actual shortfall in the income necessary to achieve the maximum possible lifestyle based on cereal consumption alone and is defined as \[CD(y(t))=V(t)-C(y(t)) =\frac{V(t) K(t)}{K(t)+y(t)}.\]
\par
 In our previous work \cite{Chattopadhyay_2010}, the parameters $V(t)$ and $K(t)$ were fixed from data that were available on a year-by-year basis from the World Bank repository \cite{NSS}. For the same reason, though, {\it the analysis there was limited to the plausible range of data without any predictive power for future times for which data is not available}.
The only way to ingrain a generic time dependence in the consumption function that can probabilistically predict values of $C(y)$ for future times is to have a time evolving model for $C(y(t))$ itself, which is the main theme of this paper, although the nature of such time dynamics may be non-unique in its character.

\section{The Inequality Model}
As a token departure from our previous theoretical analysis \cite{Chattopadhyay_2010}, this article treats  trade in commodities, other than the basic essential commodity, and of assets other than ones own personal work skills, occurs only above a certain minimum threshold level. As detailed in our earlier works \cite{Kumar_1996, Sitaramam_1996, Chattopadhyay_2007}, our choice for this minimum of income threshold is the consumption deprivation function $CD(y_i,t)$ which is the amount of income lacking in an income class $i$ that is needed to sustain trade above the poverty (or inequality) level, that, then is a function of the income $y_i(t)$ at time $t$. While the increase in the spending power of the entire population is a motivating factor and a goal for growth in income, a transfer of wealth (e.g. investment by agent $i$ through trade) will result in decreased spending power for this agent that will oppose the positive growth rate defined by ${\overline Y}_i(t)$. But such a transfer of wealth can now only happen when the income $y_i(t)$ is greater than the consumption deprivation $CD(y_i;t)$. What this implies is that a household will not have any savings in assets as long as the income is less than the consumption deprivation. 

\subsection{Stochastic income-expenditure model above threshold}
The resultant stochastic (Langevin) model that we then get from this construction is as follows

\begin{subequations}
\begin{eqnarray}
&& \frac{dy_i}{dt}  =  \beta \:{\overline Y_i}(t) - \alpha \:(y_i -CD_i) + \eta_i(t)y_i, 
\label{Langevin} \\
&& <\eta_i(t) \eta_j(t')> = D_0 \delta(t-t') \delta_{ij},
\label{Noise}
\end{eqnarray}
\end{subequations}
where $\beta {\overline Y}_i(t)$ is the maximum level of income (proportional to the mean growth rate ${\overline Y_i}(t)$ of income) at time $t$ for the ${i}^{\mathrm{th}}$ class of individuals would like to aim at but achieve only a fraction of it in each period. This is a sum of desired consumption expenditure and desired savings. This is thus partly an adaptive behavior, the first term being a motivational force that creates incentives to increase the level of income, and the second term, on the other hand, defines how increase in income is offset by a decrease in income available for consumption in the current period (due to savings). It assumes that a proportion $\alpha$ of income is saved and is not available for spending in the current period. $\eta_i(t)$ is a stochastic factor generated by the uncertain market environment; greater the value of income and trade, the larger is the stochastic disturbance to modulate the income. The contribution coming from this market uncertainty depends on the choice of distribution \cite{Metzler_2004}. Since a market consists of both assets and commodities (goods and services consumed), we can decompose income into expenditure on commodities and savings.

\par
In this article, we are primarily considering two separate markets, asset markets and commodity markets. When we deal with the commodity markets, the monetary asset allocated for consumption expenditure is taken as given, taken from the asset market solution.

Our earlier papers \cite{Chattopadhyay_2010} and \cite{Chattopadhyay_2007} dealt with the asset market equilibrium and determined the equilibrium income distribution. In this paper, we take that solution as given and examine the commodity exchange market the same way as we examined the asset market equilibrium. The basic idea of trading of commodities for consumption between agents remains more or less same as  in \cite{Chattopadhyay_2010}: an agent $i$ can spend an amount $y_i$ while being subjected to stochastic effects of trading $\eta_i(t)$ at time $t$. The rate of increase of the agent's ability to spend more money on a commodity will be principally determined by two factors, the (time dependent) maximum consumption expenditure on essential commodities by the entire reference population to which the household belongs, $V(t)$, and total income available for consumption expenditure as derived from the asset market distribution. We are thus assuming a recursive behavior, making a decision on work, income and savings first and then using the income available for consumption to make consumption expenditure decisions. 

\par
What we do now is to integrate the expenditure dynamics, including consumption deprivation, with the income distribution dynamics that is based on an Ito calculus scheme \cite{Chattopadhyay_2010}. 

\subsection{Fokker-Planck Modeling: Results and Discussions}
\label{fokker_planck}
The above form of the stochastic income-expenditure growth rate above the threshold line $CD_i(y,t)$ leads to the following Fokker-Planck equation \cite{Hanggi_2009} which depicts the time rate of change of the income distribution function

{\small
\begin{eqnarray}
\frac{\partial {\hat f}}{\partial t}(y,t) &=&  \frac{\partial}{\partial y}
\bigg\{ \bigg[(\alpha+2)y - C(t) - CD(y,t) \bigg] \:{\hat f} + \nonumber \\
&+& y^2\:\frac{\partial {\hat f}}{\partial y}\bigg\} 
\label{Fokker-Planck}
\end{eqnarray}
}

The coupled dynamics described above, involving the probability density function of income $f(y,t)$ and the consumption deprivation function $\text{CD}(y,t)$, represents the fact that effective trade only ensues when the effective mean income overshoots the actual mean $C(t)$ by a factor of $\text{CD}(y,t)$.

In the steady state ${\overline C}(t)=C_0$ and $CD(y) = \frac{V_0 K_0}{K_0+y}$ ($t \to \infty$ limit), which gives us the steady state income distribution:

\begin{equation}
{\hat f(y)}_{t\rightarrow \infty} \propto \frac{e^{-\frac{(C_0+V_0)}{y}}}{y^{\alpha+2}}\:{\left(1+\frac{K_0}{y}\right)}^{V_0/K_0},
\label{Levi}
\end{equation}
Here $V_0$ and $K_0$ are the values of $V(t)$ and $K(t)$ for the corresponding year concerned.
The proportionality constant can be evaluated from the condition $\displaystyle \int_{y_0}^\infty\:{\hat f(y)}_{t\rightarrow \infty}\:dy =1$.

\par
The parameter $\alpha$ varies from economy to economy and can be evaluated from data. For our study, we use the Indian data recorded in the Indian National Sample Survey (NSS) \cite{NSS} spanning 43 years (1959-2002) across 25 surveys and sampling about 7 million income data. Only food consumption statistics has been used, that being the modicum of basic life disbursement for the lowest income sectors (destitute). The monthly income/expense data are made available across various income sectors, popularly known as \enquote{expenditure classes}, using which the cumulative distribution function (CDF) was plotted. The CDF led to the probability density function (PDF). For the purpose of this analysis, all data have been {\it deflated} using the consumer price index (CPI) data (also available from www.worldbank.org), the conversion formula being: 

\begin{equation}
\text{Deflated\:\:expenditure}\:=\:\frac{\text{Raw\:\:expenditure\:\:data}}{\text{CPI}}.
\label{deflated}
\end{equation}

The inset shows that the IPDF emerging from our model is also 
in excellent agreement with the functional form provided in equation (\ref{Levi}).

\begin{figure}[htbp!]
\centering
\includegraphics[width=7cm]{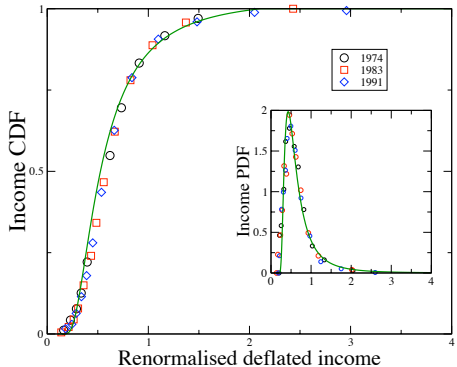}
\caption[CDF]{Plots of the cumulative distribution functions (CDFs)
against
deflated income for selected years, with inflation independently
sourced from the consumer price index (CPI) and renormalized to the
1974 mean income in rupees (64.84 INR). 
The green line is our theoretical curve, taking $y_i$ as income 
above a non-zero level below which agents would die of starvation 
(set at 0.15 in renormalized units).  
Inset shows the
IPDF which is the differential of the CDF, evaluated from
the data by interpolation.  The points are the real data from NSS, the
line is our analytic function for the steady state distribution that fits with the power-law
predicted in equation \ref{Levi} with $\alpha=1.6$.
\label{fig_cdf}}
\end{figure}

Eigenvalue analysis shows that this solution is stable against
perturbations. As shown in \cite{Chattopadhyay_2010}, equation \ref{Fokker-Planck} can be analytically solved to obtain the full time-dependent solution as a sum 
of confluent hypergeometric functions $F(a,b,z)$ 
with time-dependent coefficients:

\begin{equation}
{\hat f}(y,t) = \sum_{n=0}^{n=\infty}\:\exp(-\omega_n t)\:g_n(y) 
\end{equation}

where  $\omega_n=2\pi n$ and

\begin{eqnarray}
g_n(y) &=& B_1\:{\left ( \frac{c(t)}{y} \right )}^
{\gamma^{(1)}_{-}}\: F(\gamma^{(1)}_{-},\gamma^{(2)}_{-},-\frac{C(t)}{y}) \nonumber \\
&+& B_2\:{\left ( \frac{C(t)}{y} \right )}^
{\gamma^{(1)}_{+}}\: F(\gamma^{(1)}_{+},\gamma^{(2)}_{+},-\frac{c(t)}{y})  
\\
\gamma^{(1)}_{\pm}&=&\frac{3+\alpha \pm \sqrt{{(1+\alpha)}^2+4\omega_n}}{2}
\\
\gamma^{(2)}_{\pm}&=&1 \pm \sqrt{{(1+\alpha)}^2+4\omega_n}
\label{eigenvalue}
\end{eqnarray}

and $B_1$ and $B_2$ are constants dependent on initial conditions. For the Indian data set used, the adjustable (changing with the economy) parameter $\alpha=1.6$ that confirms fractional Brownian motion for the stochastic model \cite{Metzler_1999}.

\section{The Consumption-Deprivation Dynamics}
\label{cd_dynamics}

In the following, we demonstrate a phenomenological derivation of the \enquote{consumption deprivation} (CD) kinetics of agent $i$ who has income $y$  line at time $t$. 
This income is still deficient by $CD(y,t)$ in restoring the \enquote{consumption} function back to the saturation level $V(t)$. The question then would be the dynamical \enquote{instability} that may be created due to influx (or outflux) of wealth fraction $\Delta y$ to (or from) agent $i$. Surely, such a change will initiate a competition of two opposing forces of economics: one that will try to \enquote{neutralize} the effects of this spike in deprivation through isotropic homogenization of assets across all wealth states characterized by $y$ and the other that will strive to work against this dialectics by maximizing  a \enquote{lateral growth} leading to {\it wealth piling} in nearest income states $y \pm \Delta y$.  The first effect can be easily represented by a Laplacian diffusion term (not in money, but in CD) $\nu(t) \frac{\partial^2}{\partial y^2} CD(y,t)$ ($\nu(t)$: time dependent diffusion constant) while the latter term depends both on the {\it transient poverty} $\text{CD}(y,t)$ itself, together with that of the gradient of the poverty increase $\frac{\partial}{\partial y} \text{CD}(y,t)$ across neighboring wealth sectors $[y-\Delta y,y+\Delta y]$; in other words, on the product $\text{CD}(y,t) \frac{\partial}{\partial y} \text{CD}(y,t)$. 
This second (reflection) symmetry violating term implies that the poorest people ($y\rightarrow 0$) achieve the
fastest poverty reduction, although, such a scheme does not allow for conservation of income as is reasonable.
\par
Combining both terms respectively representing economic neutralisation around the saturation level $V(t)$ (diffusion dynamics) that is perturbed by a directed economic gradient between neighbouring agents (Burgers' nonlinearity \cite{Burgers_1974}), we arrive at our time dynamical model for consumption deprivation:

{\small
\begin{eqnarray}
&& \frac{\partial}{\partial t} CD(y,t) + CD(y,t) \frac{\partial}{\partial y} CD(y,t) = \nu(t) \frac{\partial^2}{\partial y^2} CD(y,t), \nonumber \\
&& \frac{\partial}{\partial t} CD(y,t) = V(t)K(t) 
\frac{2\nu(t)+ V(t)K(t)}{(K(t)+y)^3}
\label{Burgers_new}
\end{eqnarray}
}

The increase (or decrease) in poverty is then simply related to whether 
$\Phi(t)=2\nu(t) + V(t)K(t)>0$ (or $\Phi(t)<0$ for poverty decrease).
\par
In order to stipulate a value, or at least a regime, for $\nu(t)$, we
need to estimate the stationary state statistics of our dynamical
model with that from our previous data based studies
\cite{Chattopadhyay_2007,Chattopadhyay_2010}. As simple algebra will
suggest, a stationary state solution ($\frac{\partial}{\partial t}
CD(y,t)=0$) of our model gives $CD(y,t=t_0) = \frac{-2\nu_0}{K_0 +
  y}$, where $K_0$ is the value of the parameter $K(t)$ at $t=t_0$. A
comparison of this steady-state solution with that from
\cite{Chattopadhyay_2007} will further ensure that $\nu_0 = -\frac{V_0
  K_0}{2}$, thereby non-equivocally establishing the steady state form
of $\nu(t)$. 
For the time dynamical evolution of the CD function, we will assume a solution that is not too far from the linearly stabilized steady state defined by $\nu_0$, {\it i. e.} $\nu(t)=-\frac{V_0
  K_0}{2} + \delta \nu(t)$, ensuring time evolution of the CD-function by maintaining $\Phi(t) \neq 0$, where $\delta \nu(t)$ quantifies the non-equilibrium increment in $\nu(t)$ as a function of time. 
  
  \begin{center}
\begin{figure}[tbp]
\includegraphics[height=8.0cm,width=9.0cm]{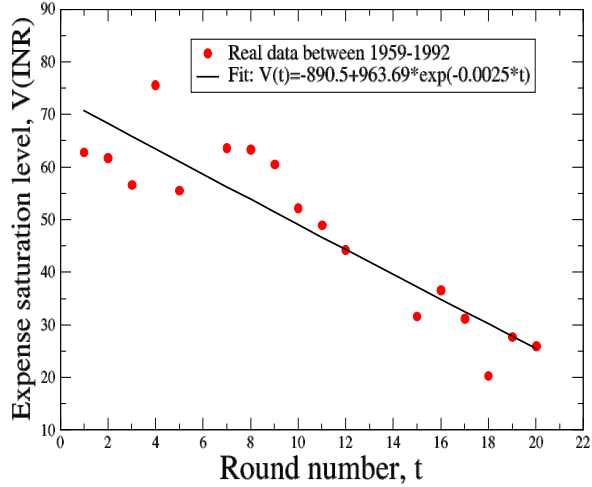}
\caption{$V(t)$ versus $t$; $V(t)=-890.5 + 963.69*e\exp(-0.0025*t)$ for $V(t)>0$. The dots represent real data points while the solid line is the least square fitted trendline through these points between $1960<t<1992$.}
\label{fig_Vplot}
\end{figure}
\end{center}

This can be done without any loss of generality since our
actual data analysis from \cite{Chattopadhyay_2010} has already
ensured that PDF statistics from each yearly data matches with the
corresponding steady state solution of the original Fokker-Planck
model given in equation (\ref{Fokker-Planck}).  The choice of such a
hydrodynamic model also ensures the implicit presence of long ranged
\enquote{hydrodynamic interactions}' that are believed to be so vital
in understanding financial peaks and troughs from agent based
modelling studies \cite{Bouchaud_book}.

\begin{center}
\begin{figure}[tbp]
\includegraphics[height=8.0cm,width=9.0cm]{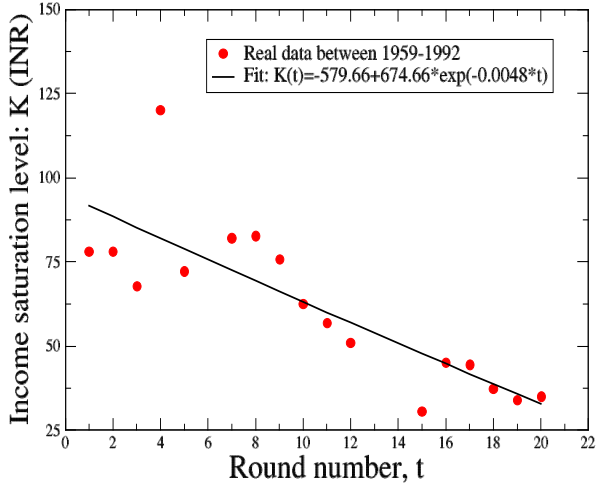}
\caption{$K(t)$ versus $t$; $K(t)=-579.66+674.66*\exp(-0.0048*t)$ for $K(t)>0$. The dots represent real data points while the solid line is the least square fitted trendline through these points between $1960<t<1992$.}
\label{fig_Kplot}
\end{figure}
\end{center}

In both figures \ref{fig_Vplot} and \ref{fig_Kplot}, the x-axes use \enquote{round numbers} depicting timelines, rather than year numbers; this is to allay the aperiodic nature of data collection over the years. 

\begin{center}
\begin{figure}[tbp]
\includegraphics[height=8.0cm,width=9.0cm]{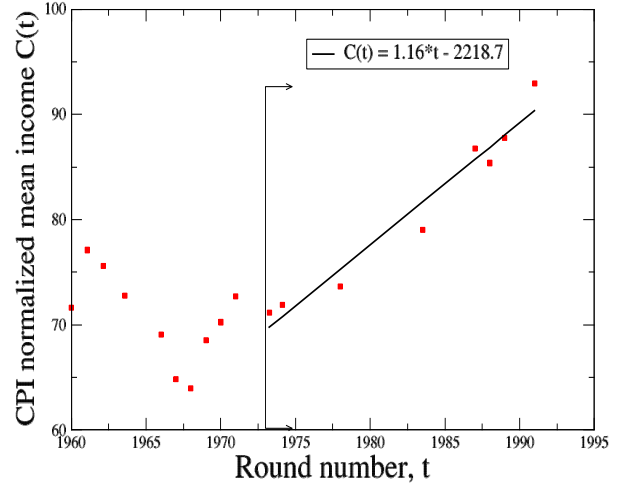}
\caption{CPI normalized mean income plotted against time. There is a monotonically increasing trend  post 1973 (indicated by a marker in the figure), the year all poverty data got standardized, a behavior that is compatible with real expectations. The corresponding least square regression fit is estimated as $C(t)=1.16*t-2218.7$ between $1973<t<1992$.}
\label{fig_Cplot}
\end{figure}
\end{center}

Using Cole-Hopf transformation \cite{Burgers_1974}, the closed form time dependent solution of the consumption deprivation function for constant parameter values $V_0$ and $K_0$ from equation (\ref{Burgers_new}) can be written as

\begin{eqnarray}
CD(y,t)  &=&  -2\nu_0 \frac{\partial}{\partial y} \log \bigg[{(4\pi \nu_0 t)}^{-1/2}  \int_{y_0}^{\infty} dy_1 \:e^{-\frac{{(y-y_1)}^2}{4\nu_0 t}} \nonumber \\
&\times& e^{-\frac{1}{2\nu_0} \displaystyle \int_{0}^{y_1} dy_2 \:CD(y_2,0)}\bigg],
\label{Cole-Hopf}
\end{eqnarray}

where $CD(y,0)$ is the initial value of the function $CD(y,t)$. This solution assumes a fixed value of the parameters $V$ and $K$ for any year as its initial condition and then uses that to arrive at CD-values for future times as given in equation (\ref{Cole-Hopf}). At the level of  linear stability analysis, this means that one does not even need to have {\it a priori }knowledge of the nature of time dependence of the variables $V(t)$ and $K(t)$ which makes the description properly probabilistic.


\section{Results and Discussions}
\label{results_and_discussions}

Both parameters $V(t)$ and $K(t)$ are known to show strong time variations (as shown in figures 3 and 4); this is a very suggestive trend and needs to be incorporated in all analyses. Figure 5 shows the variation of (CPI normalized) mean income over advancing years. The fast rising mean income in the post 1973 era, as evident from these Indian data, are commensurate with the improving economic situation of the country and is a complementary description to decreasing poverty trends, as could be seen from Figures 3 and 4 respectively.  Figure 6 reminds one of the Engel prediction \cite{Engel_1895} that as $y$ increases, the $C(y)/y$ fraction linearly converges to a very low value (data from 1989, as in Figure 1).

In the following, we solve the dynamical system of equations (\ref{Fokker-Planck}), (\ref{Burgers_new}) and (\ref{poverty_index}) numerically for the initial value $CD(y_0) = \frac{V_0 K_0}{K_0 + y_0} \approx \frac{73.19 \times 95}{95 + 30} = 55.62$ using nonlinear regression fits from real data ($V(t) = -890.5+963.69 \exp(-0.0025*t)$ and $K(t)=-579.66+674.66*\exp(-0.0048*t)$; $V(t),\:K(t)\:>0$) as shown in figures 3 and 4 to obtain the time series estimate of the poverty (Figure 7). In the limit of destitution ($y\to0$), one can easily see that $C(y)/y(t)\to\dfrac{V(t)}{K(t)}$, that is the budget share of essential commodities. The quantity $V_0/K_0$ (= 0.77)  is the budget share of the poor. The proposed formulation of the CD-dynamics thus has far reaching consequences, in that it reconfirms the Engel prediction in the steady state limit \bigg($\dfrac{\partial}{\partial t}$CD = 0\bigg), the fact that $C_{\text{steady state}}(y) =\dfrac{V_0 \:y}{K_0 + y}$, where $V_0$ (=73.19) and $K_0$ (=95) are the fixed point values of $V(t)$ and $K(t)$ respectively. The \enquote{theoretical turnover time} defining the
validity of the nonlinear regression fits of $V(t)$ and $K(t)$ are accurate up to about 31.5 rounds, that in year numbers is roughly equal to 2005, as shown in Figure 7. The model is not restricted to these numbers, though, since a running average can be continuously done to
stretch the range of validity to whatever timeline is required.

\begin{center}
\begin{figure}[tbp]
\includegraphics[height=8.0cm,width=9cm]{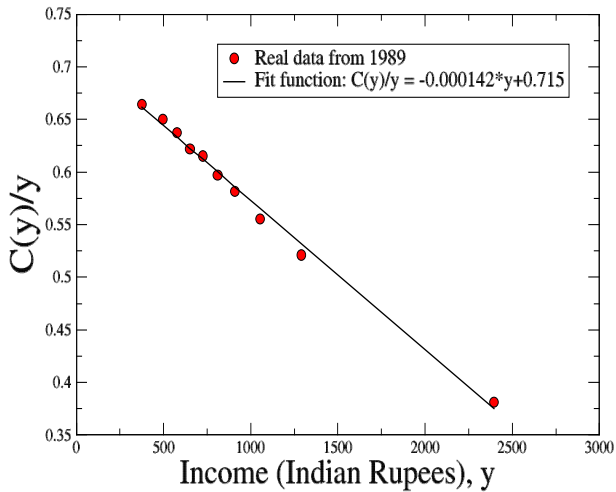}
\caption{A plot of the ratio of the consumption function to the income $C(y)/y$ that is seen to follow a linear decaying trend with $y$ (data as in Figure \ref{fig_consumption}). 
\label{fig_ratio}}
\end{figure}
\end{center}

\begin{center}
\begin{figure}[tbp]
\includegraphics[height=8.0cm,width=9.0cm]{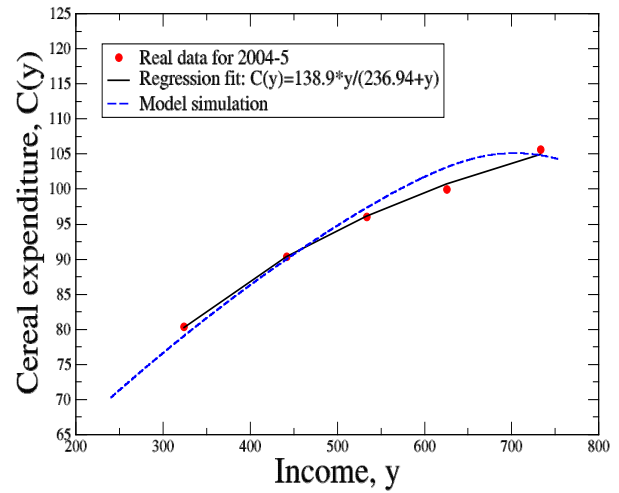}
\caption{Plots comparing cereal expenditure versus income (all in Indian rupees) data for the year 2004-5 \cite{suryanarayana} (solid circles) with the Engel form fit (solid straight line) and the numerical solution (dots) of equation (\ref{Burgers_new}).
\label{fig_burgersy}}
\end{figure}
\end{center}

Figure \ref{fig_Cplot} shows the variation of (CPI normalized) mean income over advancing years. The fast rising mean income in the post 1973 era, as evident from these Indian data, are commensurate with the improving economic situation of the country and is a complementary description to decreasing poverty trends, as could be seen from Figures \ref{fig_Vplot} and \ref{fig_Kplot} respectively.

The dotted points in Figure \ref{fig_ratio} are the actual data points that are compared against a regression fit straight line. This linear plot reminds one of the Engel prediction \cite{Engel_1895} which originally explained why food expense related expenditure can only follow a \enquote{monotonically decaying} profile with increasing income. The statistics in Figure \ref{fig_burgersy} complements this description.

As a major confirmation of the hypothesis used to track the CD-dynamics, Figure \ref{fig_burgersy} shows a comparison of the actual model data \cite{suryanarayana} (dotted line) for the year 2004-5 with the Engel curve (solid line) hypothesis and thereafter, an income:expenditure analyzed from a solution of equation (\ref{Burgers_new}). The simulation here uses the deterministic limit of equation 2, in that this leads to an average income that can also be calculated from equation (3) using the relation $<y>=\displaystyle \int_{y_0}^{\infty}dy\:y {\hat f}(y,t)$. In this estimation, we have used the functional representation of $V(t)$ and $K(t)$ as shown in Figures 3 and 4. The $y$ shown in this figure is actually this ensemble averaged $<y>$ and the corresponding consumption function is $C(<y>)$; for brevity as also for the sake of our general line of reasoning, we have dropped the curly brackets \enquote{$<>$} in the plot. The quality of Engel's predictions and the strength of our complementary model can be justified both from the fit (solid line against real data circles) as well as from the solution (dotted line) of our proposed model represented in equation (\ref{Burgers_new}).

As always, the strength of an assumption can only be proved from the efficacy of its quantitative measurable output, that, in our case, will be the \enquote{poverty index}. Following \cite{Atkinson_2011_1, Atkinson_2011_2, Kumar_1996, Chattopadhyay_2010}, the poverty index $P_{CD}(t)$ is defined as the statistical average of the consumption deprivation function across the entire range of income:

\begin{equation}
P_{CD}(t)  =  \int_{y_0}^{\infty} dy\:CD(y,t) \hat f(y,t) \label{poverty_index} 
\end{equation}

where $CD(y,t)$ and $\hat f(y,t)$ will be respectively obtained from equations (\ref{Burgers_new}) and (\ref{Fokker-Planck}). The time varying mean income $C(t)$ present in equation (\ref{Burgers_new}) can be evaluated from the relation $C(t)=\displaystyle \int_{y_0}^{\infty}dy\:y f(y,t)$ which irons out the oscillatory instability that would otherwise crop up in the numerical simulation should the $C(t)=1.16*t-2218.7$ fit function be used instead.


\par
Remarkably, from our (deflated) NSS data over 23 years, it can be seen from Figs. \ref{fig_Vplot} and \ref{fig_Kplot} that both parameters admit of approximate linear regression fits with time decaying trends. In the post 1973 regime, when all poverty data were renormalized for the first time, the mean income profile too follows an upward linear trend as shown in Fig. \ref{fig_Cplot} that fits well with the decaying linear trends depicted in the $V(t)$ and $K(t)$ regression fits. The legends to these figures show the relevant extrapolation formulae ($V,K>0$, $\forall\:t$). Apart from clearly indicating a (linear) trend in the deflated time series statistics, the starting points (round 6) of both these plots define the initial condition $CD(y,0)$ that is needed to solve equation (\ref{Fokker-Planck}). The progressively diminishing values both for $V(t)$ and $K(t)$ are not so difficult to anticipate, since the It must be noted, though, that such fits are limited to positive definite values of $V$ and $K$ only.


As a cross-check of the strength of our theory, we compared our new theoretical 
poverty index with data from all three indices popularly used in the literature: the head count (HI) index, the poverty gap (PG) index and the squared poverty gap (SPG) index. Additionally, we also compared this with a similar statistics obtained from our previous work (Figure 4 in \cite{Chattopadhyay_2010}). 
\begin{center}
\begin{figure}[tbp]
\includegraphics[height=8.0cm,width=9.0cm]{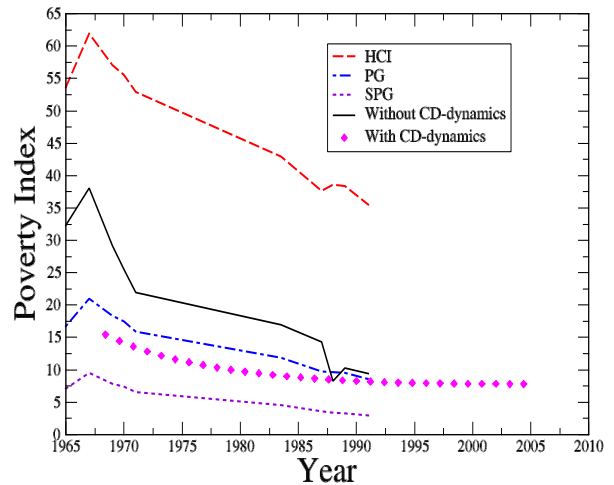}
\caption{Plots of the different poverty indices - head-count-index (HCI; dashed), Poverty Gap (PG; dotdashed) index, Squared Poverty Gap (SPG; dotted) index - against the previous \cite{Chattopadhyay_2010} 
work ("Without CD-dynamics": solid line) and that against the new CD-dynamics ("With CD-dynamics": diamonds) adjusted estimation. Post 1965, the mean profile of the CD-dynamics modified theoretical index is shown to be in a steady decline across the years. The new theoretical index has also the major advantage of being able to forecast future poverty line statistics based on linear regression fits as depicted in Figures \ref{fig_Vplot} and \ref{fig_Kplot}. The part beyond 1992 shows the prediction made from this theory that can be tested against real data.
\label{fig_povI}}
\end{figure}
\end{center}
\par
The result (Fig \ref{fig_povI}) shows decent agreement between the new CD-dynamics modified index with all three indices compared to that in \cite{Chattopadhyay_2007}, a plot that is represented in Fig \ref{fig_povI} as \enquote{CD-fit theory}. While the new time dynamics adjusted poverty index does not improve the quantitative result compared to the poverty index calculated earlier \cite{Chattopadhyay_2010}, its strength lies in its probabilistic prediction ability as is demonstrated in the values beyond 1992. An important stabilizing feature of this mechanism is that the evaluation of this new CD-dependent poverty index does not rely on any regression fitting of the time evolution of the mean income of the population. This arguably edges out the occasional hikes (as in 1967) and dips (as in 1988) but at a statistical level, this gives quite reliable estimates of the poverty index for future times. The strong upward trend shown in the mean income variation with time (Fig. \ref{fig_Cplot}) is ingrained in this analysis through the usage of the ensemble averaged definition $C(t)=\displaystyle \int_{y_0}^\infty\:y {\hat f}(y,t)\:dy$ and should be the definition used; this obfuscates the strong oscillations shown in the growth profile of $C(t)$. In our simulations, we considered $y_0=30$ as the CPI adjusted per capita daily income.

In order to cross check the veracity of this modified poverty index, we used HCI data from \url{www.worldbank.org} for the years 2005 and 2010, together with relevant CPI values from the Reserve Bank of India (\url{www.rbi.org.in}) for these two years. The relative HCI indices for these two years (37.2\% in 2005 versus 29.8\% in 2010) scale identically as the new model index (2.16 units in 2005 against 1.73 units in 2010) to within 95\% accuracy.

Such basic success in probabilistic prediction of the gross qualitative aspects of the poverty index evolution emphasises the need for advancing more accurate models based on more elaborate data analysis that will go beyond the present restrictions through more accurate prediction of the primary agent based Langevin model (\ref{Langevin}) and thereafter integrating the same with the presently espoused consumption deprivation dynamics.
The dynamical equation (\ref{Langevin})
also gives some indication of the effect of
redistributed income between expenditure classes on overall mean income. Typically we find that
such redistribution reduces poverty during the period of application,
but suppresses the mean income, such that if the redistribution is
removed (or, in the long term, even if it remains) then the poverty
may increase again, unless some external agent, such as improved
technology, contributes in a still faster rate of increase of the mean income.

\section{Conclusions}
\label{conclusions}

The result demonstrated in Figure 8 shows decent agreement between the new CD-dynamics modified index with the HCI, PG and SPG indices, together with the theoretical index (constrained to data only) previously estimated in \cite{Chattopadhyay_2007}, a plot that is represented in Figure 8 as \enquote{CD-fit theory}. While the new time dynamics adjusted poverty index does not improve the quantitative result compared to the poverty index calculated earlier \cite{Chattopadhyay_2010} and hence is a \enquote{null result} from that context, the uniqueness of this new study is in its ability to predict the future indices. The fact that parameters $V(t)$ and $K(t)$ are now regressively linked to the core dynamics of the poverty evolution, this new model allows for probabilistic future predictions. In order to cross check the veracity of this modified poverty index, we used HCI data from \url{www.worldbank.org} for the years 2005 and 2010, together with relevant CPI values from the Reserve Bank of India (\url{www.rbi.org.in}) for these two years. The relative HCI indices for these two years (37.2\% in 2005 versus 29.8\% in 2010) scale identically as the new model index (2.16 units in 2005 against 1.73 units in 2010) to within 95\% accuracy. At this point, it is worthwhile to remind that since we are rejecting the focus axiom of Sen ({\it i. e.} the poverty line) \cite{Sen_1976}, strictly speaking, there is no point in validating our results with HCI and PGI that depend on that axiom. But as both indices track consumption deprivation, one with a poverty line and the other without  (in the traditional poverty economics these indices are referred to as exclusive and inclusive indices), we only check for the direction of trend and not the actual magnitude. 

Such basic success in probabilistic prediction of the gross qualitative aspects of the poverty index evolution emphasises the need for advancing more accurate models based on more elaborate data analysis that will go beyond the present restrictions through more accurate prediction of the primary agent based Langevin model (\ref{Langevin}) and thereafter integrating the same with the presently espoused consumption deprivation dynamics. The dynamical equation (\ref{Langevin}) also gives some indication of the effect of
redistributed income between expenditure classes on overall mean income. Typically we find that
such redistribution reduces poverty during the period of application,
but suppresses the mean income, such that if the redistribution is
removed (or, in the long term, even if it remains) then the poverty
may increase again, unless some external agent, such as improved
technology, contributes in a still faster rate of increase of the mean income.

\par
In summary, we have generalized the scopes of our earlier work on alternative and self-consistent poverty index calculations based on mathematical models \cite{Chattopadhyay_2010}, now to include probabilistic measurements of the Inequality Index above the minimum income threshold line. The data used in our analysis were confined to years 1959-1992 but using relatively recent data from years 1993 and 2004, two not-so-close years where the fiscal dynamics are expected to be relatively independent of each other, results obtained from our model compare favorably with the representative HCI indices (\url{www.worldbank.org}) as would be evident from the approximately parallel lines in Figure \ref{fig_povI}. Beyond 1992, though, our model predicts a much slower precipitation of the Poverty Index (PI) compared to the conventional head count index, eventually reaching an effective plateau at around year 2004-5; we take this as a clear signature of a fast stabilizing economy (India in this case) that seems to be justified from analysis of recent data for the year 2004-5 \cite{suryanarayana}. The combination of results from numerical solution of equation (\ref{Burgers_new}) as shown in Figure 7 and the Engel fit form $C(y)=\dfrac{V\:y}{K+y}$ for 2004-5 data \cite{suryanarayana} proves that the corresponding income needed to consume an amount $V/2$ (= 69.5) is matched by a corresponding saturation income of $y=K$ (= 236.94). In other words, our model is robust enough to forecast an effective \enquote{steady state} in the poverty dynamics of the Indian economy at around 2004-5 that shows up as a plateau in the PI time profile. This result is a major improvement over the HCI index based conventional studies which fail to detect this steady state. Such a confirmation demonstrates the strength of this new model based approach; this also emphasizes the need for such alternate poverty measures over the conventional HCI/PG/SPG based poverty evaluation. While in the absence of any dynamic models for the dominant parameters ($V$ and $K$), predictions can never be \enquote{full proof}, the present approach transcends all previous modeling results in the robustness of its prediction accuracy. Our model is based only on cereal consumption. The data capture from the real world data would be hopefully much better if the model is based on a multi-market trade equilibrium using consumption expenditure data on several commodities. Current works are in progress to enhance the scopes of the theoretical model even further by including non-essential foods and non-food grain statistics in a multivariate structure to arrive at an even more robust description of the poverty index.

\section{Acknowledgments}
\label{acknowledgments}

AKC acknowledges partial research support from Royal Society grant number RSO11137. Discussions with G. J. Ackland and Ewa Grela are thankfully acknowledged. TKK acknowledges his debt to the coauthors of his earlier studies on Engel curve based poverty measure, Sitaramam and Gore.

\end{document}